# Velocity map imaging of inelastic and elastic low energy electron scattering in organic nanoparticles

O. Kostko,[1, a)] M. I. Jacobs,[1,2] B. Xu,[1] K. R. Wilson,[1] and M. Ahmed[1]

[1] Chemical Sciences Division, Lawrence Berkeley National Laboratory, Berkeley, CA 94720 (USA)

[2] Department of Chemistry, University of California, Berkeley, CA 94720 (USA)

a) Electronic mail: OKostko@lbl.gov

**ABSTRACT**

Electron transport is of fundamental importance, and has application in a variety of fields. Different scattering mechanisms affect electron transport in solids. It is important to comprehensively understand these mechanisms and their scattering cross-sections to predict electron transport properties. Whereas electron transport is well understood for high kinetic energy (KE) electrons, there are inconsistencies in the low KE regime. In this work, velocity map imaging soft X-ray photoelectron spectroscopy is applied to unsupported organic nanoparticles to extract experimental values of inelastic and elastic mean free paths. The obtained data is used to calculate corresponding scattering cross-sections. The data demonstrates a decrease of the Inelastic Mean Free Path and increase of the Elastic Mean Free Path with increasing electron KE between 10-50 eV.

**INTRODUCTION**

Electron transport in condensed matter is a fundamental problem in physics and has applications in a variety of fields, such as radiation biology, oncology, astrochemistry, materials design, and transport.[1–3] Furthermore, electron interactions with condensed matter have allowed analysis methods such as electron microscopies, X-ray photoelectron and electron energy loss spectroscopies to become exquisite tools to probe nanoscale physical, chemical, and biological processes. While electron interactions with kinetic energy (KE) above 50 eV in the condensed phase are relatively well understood, there is a paucity of both experimental and theoretical information at lower KE. For example, in radiation biology, low energy electrons and their interactions with water and biological molecules can give rise to a myriad of damage processes (e.g. DNA mutations) which are not well understood.[1] In extreme ultraviolet lithography, emitted electrons (primary and secondary) and photons both, initiate chemical reactions. The mean free paths of the electrons inside a resist film is intimately connected to the resulting patterning resolution which is critical to the coming microelectronics revolution.[4]

Electron interactions with condensed-phase species can be separated in two general types: elastic and inelastic scattering. Elastic scattering of electrons by the Coulomb potential of a nucleus changes their trajectories but does not affect their kinetic energies. Inelastic collisions generally lead to a reduction of electron KE. There are several mechanisms responsible for the inelastic scattering of low KE electrons: phonon excitation, electron attachment, intra- and interband excitations, including plasmon excitations and electron impact ionization. Some of these processes are schematically depicted in Fig. 1. Corresponding elastic and inelastic mean free paths (EMFP and IMFP) and scattering cross-sections are used to quantitatively characterize these processes.

However, it is very challenging to experimentally determine IMFP and EMFP, especially in the low KE regime where elastic scattering becomes more important. The IMFP is the mean distance travelled by an electron of a particular KE between inelastic scattering events. When elastic scattering is insignificant, an electron moves linearly and the IMFP can be determined experimentally by methods such as substrate-overlayer technique or low energy electron transmission (LEET).[2] In LEET, films of varying thickness are exposed to an electron beam, and the current generated by electron transmission through the film can provide information on electron attenuation length (EAL). The EAL is the film thickness that results in a 1/e decrease in signal intensity and is roughly equal to the IMFP in absence of elastic scattering. In the substrate-overlayer technique, electrons are generated in a substrate using photoemission. The transmission of photoelectrons through films on top of the substrate is used to measure EALs. When elastic scattering becomes strong, it significantly affects electron trajectories and decreases the EAL in comparison to the IMFP. The difference between them can ultimately reach 30%,[5] making these experimental techniques unsuitable for low KE IMFP determination. Although knowledge of the EAL has important practical applications, e.g. depth profiling in X-Ray photoelectron spectroscopy (XPS), only the IMFP and EMFP provide fundamental information on electron transport.

While the methods described above work well for EAL determination in solid samples, technical difficulties arise when they are applied to liquid samples. For example, one obstacle is the inability to create a layer of liquid of known thickness. Also, photoemission experiments on liquids and aerosols are non-trivial compared to solids, because liquids, such as water, have high vapor pressures, which can affect operation of detection electronics. High vapor pressures also lead to large gas-phase backgrounds which can scatter emitted electrons and change their energy and/or direction.[6,7] Furthermore, in condensed samples, strong elastic scattering may significantly change the initial angular distribution of photoelectrons, substantially complicating analysis of low KE electrons. Therefore, the photoelectron angular distribution (PAD) needs to be considered when elastic and inelastic interactions of low KE electrons are studied.

There have been a limited number of studies which have measured low kinetic energy electron scattering in liquids. Trümer et al. used PAD of O 1s electrons emitted from a liquid jet to examine the IMFP in water. The change of PAD from bulk water with respect to that from gas-phase water molecules revealed information on the IMFP/EMFP ratio.[8] This ratio (equal to the average number of elastic collisions before an electron inelastically scatters), coupled with their previous work, where the PAD was approximated to be energy independent,[9] led to correct EAL values for liquid water (KE = 25 – 1000 eV). A similar approach was used to study elastic electron scattering in $SiO_2$ nanoparticles using soft X-ray photoelectron spectroscopy by Antonsson et al. to extract the number of elastic collisions for electrons with KE = 20 – 300 eV.[10] Suzuki et al. combined their own experimental photoemission data with the literature data to extract EAL in liquid water in the 10 – 600 eV region.[11] Signorell et al. were able to extract IMFP and EAL for 1-3 eV KE electrons in water by employing two-photon UV ionization coupled with velocity map imaging (VMI) spectrometry and modelling the scattering events.[12] Our group used the substrate-overlayer method applied to nanoparticles to investigate EAL in organic layer using VUV radiation to generate photoelectrons in the nanoparticle inorganic core.[13] This study was complicated by the wide energy distribution of emitted valence electrons.

Recently we developed a VMI photoelectron spectrometer, capable of collecting electrons with KE up to 100 eV and performed X-ray photoemission experiments on unsupported nanoparticles. The VMI technique applied to unsupported nanoparticles has a number of advantages over conventional photoelectron spectroscopy techniques (e.g. hemispherical electron energy analyzers). VMI collects the full 4π steradian distribution of emitted electrons. It is also capable of collecting low KE secondary electrons as well as providing information on the photoelectron angular distribution from a single image. The use of spherically symmetrical nanoparticles allows for emission of electrons from any side of the nanoparticle with respect to light direction and does not affect any of experimental observables (e.g. PAD). We also demonstrated, that the signal of the low KE secondary electrons, collected by VMI, can be used to perform Near Edge X-ray Absorption Fine Structure (NEXAFS) spectroscopy of unsupported nanoparticles, including aqueous aerosols.[14]

In this work, soft X-ray photoemission measured with a VMI spectrometer is used to explore inelastic and elastic scattering of electrons in condensed medium, represented by unsupported liquid branched hydrocarbon nanoparticles. Core-level carbon electrons are probed by single photon X-ray ionization. The elastic and inelastic mean free paths of photoelectrons as well as corresponding electron scattering cross-sections for low KE electrons are extracted using the narrow primary photoelectron peak, its PAD, and secondary electron emission intensity. We discuss the application of the values obtained for depth profiling in X-ray photoelectron spectroscopy.

**EXPERIMENTAL METHODS**

Photoelectron spectroscopy of squalene ($C_{30}H_{50}$, molecular structure is shown in Fig. 2a) nanoparticles was performed using a VMI photoelectron spectrometer, which has been described in detail elsewhere.[14] Nanoparticles of squalene were formed via homogeneous nucleation by passing 0.4 LPM of dry nitrogen over a 155 °C heated reservoir containing pure squalene. As the squalene vapor cooled exiting the reservoir, particles nucleated into sizes that were log-normal in distribution with an average diameter of ~220±40 nm. The size distribution as well as stability of the nanoparticle flow was monitored using a scanning mobility particle sizer (SMPS). The squalene nanoparticles entered the VMI spectrometer through a 200 µm nozzle and an aerodynamic lens (ADL). The ADL focused the nanoparticles into a ~100 µm diameter beam, which passed through two stages of differential pumping and intersected focused X-ray radiation orthogonally in the center of the VMI ion optics. All X-ray measurements were collected at beamline 11.0.2 of the Advanced Light Source at the Lawrence Berkeley National Laboratory. We estimate that during the experiment, the photon beam only interacts with 6-7 nanoparticles at any one time.[13]

In the VMI spectrometer, a projection of the nascent photoelectron distribution is velocity mapped onto an imaging detector (consisting of multichannel plate and phosphor screen) and imaged using a CMOS camera. Background images were collected by removing the nanoparticles with a HEPA particle filter so that only gas phase species entered the VMI spectrometer. The background images were subtracted from those of unfiltered nanoparticles, and the resulting images were analyzed using pBASEX code to extract a photoemission spectrum.[16] Squalene has very low vapor pressure of $3 \cdot 10^{-7}$ Pa at room temperature. Therefore, evaporation from the nanoparticle surface was negligible such that there was no detectable gas-phase contribution to the photoelectron spectra.[17] X-ray photon flux was measured using an SXUV-100 photodiode. The VMI spectrometer was energy calibrated using nitrogen K edge photoemission from gaseous $N_2$.

**RESULTS AND DISCUSSION**

The section is organized as follows. In a) we describe how information is extracted from experimental VMI photoelectron spectra and how the primary and secondary electron signal is used to obtain IMFP. In b) Monte Carlo simulations together with experimental PAD values are used to determine EMFP. The obtained IMFP and EMFP values are then used in c) to determine the corresponding scattering cross-sections. And finally, in d) the results are discussed in terms of applicability to perform depth profiling experiments using low KE electrons.

a) Inelastic scattering of electrons

An example of a velocity mapped image from squalene nanoparticles irradiated by 315 eV photons is presented in figure 2a. Only one half of the raw image (reflection symmetrical with respect to vertical line) is shown in Fig. 2a. The other half of the image corresponds to the reconstructed image using the pBASEX algorithm. The reconstructed image reveals a thin line, corresponding to emission of primary C 1s photoelectrons, and a diffuse background in the center of the image, corresponding to emission of low KE secondary electrons. The intensity of C 1s signal is not isotropic; the signal is more intense along the polarization axis of the X-ray beam. This observed anisotropy of the primary electron emission arises from the angular distribution of the emitted photoelectrons and will be discussed later.

Several photoelectron spectra, obtained from the reconstructed VMI images and collected at different photon energies are presented in Fig. 2b. Spectral intensities are normalized to photon flux to facilitate direct comparison. The KE of C 1s photoelectrons (intense narrow peaks) increases with photon energy. The low KE background has a peak at ~3 eV and arises from emission of secondary electrons from the squalene nanoparticle. The low KE secondary electrons emerge after inelastic scattering of either a photo- or, more likely, an Auger-electron, produced from the C 1s hole decay. As discussed previously,[14,15] this secondary electron signal measured as function of photon energy can be used to record a nanoparticle X-ray absorption spectrum (XAS). An example XAS spectrum of squalene nanoparticles is shown in Fig. 3a. The spectrum has a sharp peak at 284.6 eV due to the carbon 1s → π* transition and a broader feature, starting at ~287 eV, which is due to transition of 1s electrons to the continuum of states. The black line in Fig. 3a is obtained from the secondary electron intensity (extracted from data shown in Fig. 2b) normalized to the photon flux.

The intensity of the total electron yield in XAS spectroscopy[18] is proportional to the number density (of carbon atoms in this study) $n$, the X-ray absorption cross-section $\sigma(h\nu)$ and the photon energy $h\nu$:

$$I_{XAS}(h\nu) \sim n\, \sigma(h\nu)\, h\nu. \qquad (1)$$

Equation (1) allows one to extract the real photoabsorption cross-section of squalene from the XAS spectrum. The extracted values are measured experimentally from the squalene droplets and are not limited to the "atomic-like" semiempirical model,[19] often used to estimate unknown absorption cross-sections. A product of semiempirical photoabsorption cross-section of $C_{30}H_{50}$[19] and photon energy is shown in Fig. 3a by a blue line for comparison. The line intensity is scaled to fit the experimental data at the photon energy range 330 – 340 eV, which is about 40 – 50 eV above the C 1s edge. Below this energy, the XAS spectra of molecular compounds may have $\sigma^*$ shape resonances,[18] seen here quite clearly between 288 – 310 eV. Because of this shape resonance, the experimental spectrum deviates from the purely "atomic-like" model depicted by the blue line in Fig. 3a. The presented data clearly demonstrates that for analysis of low KE electrons, a knowledge of the experimental photoabsorption cross-section is required. Substitution of those values by a semi-empirical model may lead to large uncertainties.

Intensities of the C 1s photoelectron peak were extracted after subtracting the secondary electron background from the spectra presented in Fig. 2b. The KE dependence of the photoelectron peak area is presented in Fig. 3b. At the KE above 10 eV, the peak area decreases exponentially. While photoelectron peaks are observed for Ke ≤ 10 eV, the peak intensities were affected by the strong secondary electron background. Therefore, peak areas in this range (KE <10 eV) are excluded from the analysis.

The experimental intensity of a C 1s signal in X-ray photoelectron spectroscopy is given by:[20]

$$I_{XPS}(h\nu) \sim n\, \sigma(h\nu)\, \lambda_i(h\nu), \qquad (2)$$

where $\lambda_i(h\nu)$ is the energy dependent IMFP of a photoelectron. Although there is a discussion in literature[5] of whether IMFP or EAL is more correct for characterizing the photoelectron, we believe the IMFP is the correct parameter to use here. This is because only those photoelectrons, which did not scatter inelastically during their travel inside of the nanoparticle comprise the photoelectron peak detected in the experiment.

Combining Eq. (1) and (2) eliminates the number density $n$ and the absorption cross-section $\sigma(h\nu)$ to provide the value of the IMFP from the experimental data:

$$\lambda_i(h\nu) = A\, I_{XPS}(h\nu)/\, I_{XAS}(h\nu)\, h\nu, \qquad (3)$$

where $A$ is a proportionality coefficient. Equation (3) depends only on the values of photo- and secondary electron yields, $I_{XPS}(h\nu)$ and $I_{XAS}(h\nu)$. Both values are experimentally measured and can be used to provide relative values of the IMFP. However, to obtain absolute values of the IMFP, an unknown coefficient $A$ needs to be determined by comparing at least one relative IMFP obtained from Eq. (3) to an absolute value of the IMFP from the literature.

Seah and Dench introduced a "universal curve"[21] in an early compilation of the IMFP values for organic compounds. However, most of the experimental data has been collected at high electron KE and only several data points have been obtained at KE below 100 eV. The transport of electrons in polymers was studied using the substrate-overlayer technique, in which the organic overlayer thickness was changed to determine the attenuation length of the emitted electron.[22–24] The electron attenuation length in paraffin $n$-$C_{36}H_{74}$ was measured to be ~2 nm for 50 eV electrons.[24] A similar overlayer approach was used by Graber et al. to study electron attenuation length in monolayers of aromatic molecules PTCDA (perylene-3,4,9,10-tetracarboxylic dianhydride).[25] The authors analyzed decay of emission from a silver substrate, as well as increase of intensity of the carbon line with increasing number of PTCDA monolayers. They found an EAL of approximately 0.4-1.0 nm for electrons with KE of 50 eV, the range constrained with large error bars and scattering of the experimental data. Ozawa et al. reported the EAL for 50 eV KE electrons to be 1.6 nm for a π-conjugated organic semiconductor material, 2,2',2''-(1,3,5-benzinetriyl)-tris(1-phenyl-1-H-benzimidazole) using the overlayer approach.[26]

Using calculations based on experimental optical data, Tanuma et al.[27] compiled IMFP values for 14 organic compounds. The minimum electron KE for which their calculations are well constrained is ~50 eV above the Fermi level. At this energy, the minimum and maximum IMFPs for the compounds studied were 0.53 and 0.78 nm for polyacetylene and PMMA, respectively. The average IMFP was 0.68 ± 0.06 nm. Out of the 14 compounds covered in that paper, 26-n-paraffin is the most chemically similar to squalene. Thus, we use the IMFP of 0.70 nm for 26-n-paraffin to place our relative IMFP values on an absolute scale.

The absolute values of the IMFP obtained using Equation (3) from the experimental data are shown in figure 4a. The data are scaled to 0.70 nm for 50 eV KE electrons, as discussed above. The IMFP reaches its maximum value of 1.6 nm at 11.8 eV KE. For the range of KE measured here, the IMFP decreases exponentially with increasing KE. For comparison, Fig. 4a contains a compilation of literature values of IMFP or EAL for different organic materials. The universal curve reported by Seah and Dench has as minimum value for 20 eV KE,[21] whereas the IMFP values, obtained in this study are still decreasing for 50 eV KE. Better correlation of the current data is observed with values of the EAL obtained by Ozawa et al.[26] and the IMFP obtained by Tanuma et al.[27] The EAL values obtained by Graber et al. are in good agreement with our experimental IMPF at 40 eV KE, but they do not show the pronounced KE dependence in this low energy range.[25] The EAL data of Cartier and Pfluger[23] shows initial growth of EAL with increase of KE which is qualitatively explained as increasing longitudinal optical (LO) phonon excitation at low KE. This excitation is associated with the stretching modes of the –$CH_3$ and –$CH_2$– units and leads to optical phonon emission (Fröhlich scattering). With increasing electron KE, the LO-photon scattering decreases and leads to larger EALs. At higher KE energies, acoustic phonon emission becomes important. The maximum EAL occurs where neither LO nor acoustic phonon scattering is efficient.[23,24] At higher electron KE (above the band gap energy) electron impact ionization becomes the main inelastic scattering channel.

In our previous work, a substrate-overlayer approach was used to determine low KE EAL for squalane (long chain hydrocarbon, $C_{30}H_{62}$, the data are shown in Fig. 4 by gray circles) similar to squalene.[13] The data show a growth of EAL with decreasing electron KE in accord with the "universal curve" hypothesis. A thick gray line is used in Fig. 4 to outline a tentative variation of IMFP (or EAL) with change of electron KE. The line is shown only as a guide to the eye and as will be discussed in d) one cannot compare directly IMFP and EAL in the low KE regime.

b) **Elastic scattering of electrons**

In Fig. 2a, we observe an anisotropy of C 1s photoelectron intensity due to the preferential emission of 1s electrons along the polarization direction of the X-rays. For an isolated atom ionized by linearly polarized light the angular distribution of the photoelectron differential cross-section is described by:[28]

$$\frac{d\sigma}{d\Omega}(h\nu, \Theta) = \frac{\sigma_{total}(h\nu)}{4\pi}[1 + \beta(h\nu)(3cos^2\Theta - 1)/2], \quad (4)$$

where $\sigma_{total}(h\nu)$ is the total photoabsorption cross-section and β is an asymmetry parameter, which can have values from -1 to 2 (β = 0 corresponds to an isotropic angular distribution of photoelectrons). Emission of an electron from an atomic *s* orbital results in an outgoing *p*-wave electron, polarized in the same direction as the light source, corresponding to photon energy independent β = 2. However, the formula was developed to describe photoemission from atoms and even for the randomly oriented diatomic molecules, the asymmetry parameter becomes energy dependent and deviates from β = 2 for K-shell photoemission. This originated from the interaction of an emitted electron with the anisotropic molecular field, an effect known as a shape resonance.[29] The same effect leads to the enhanced photoabsorption cross-section for squalene above the C 1s edge, as observed in Fig. 3a.

Values of the asymmetry parameter β extracted from the experimental VMI spectra using pBASEX software are presented in Fig. 5 by black circles. At KE >10 eV, the β parameter increases linearly, reaching a value of 1.18 for KE of 41.7 eV. For comparison, Figure 5 also shows experimental and theoretical data for gas-phase CO and $C_{60}$ molecules.[30–32] The gas-phase data demonstrate significantly different values for the asymmetry parameter. Both theoretical and experimental data demonstrate a minimum β value of ~0.4 at KE ~7 eV and ~5 eV for CO and $C_{60}$, respectively. After the minimum, the β parameters of gas-phase species increase with electron KE towards the maximum value β = 2. The β parameters reach a value of 1.6 at 20 eV KE; at KE >20 eV, the rise of gas-phase β values are slow and saturate at value of 1.8 for 60 eV KE. This energy dependence of β is can be explained either by a shape resonance for the CO molecule or by elastic scattering of the photoelectron by the atoms in the large $C_{60}$ molecule.[31–33]

In the case of condensed squalene nanoparticles, two independent factors affecting the angular distribution of detected electrons need to be considered: 1) photoemission and 2) elastic scattering. Firstly, after absorption of a photon, a C 1s electron might be emitted. The angular distribution of the emitted electron is described by formula (4) with a small correction to account for shape resonances. The emitted photoelectron, before being detected, might travel 0.7 – 1.6 nm (i.e. the IMFP) from its point of origin to the surface to escape from a nanoparticle. The inelastically scattered electrons will change their KE and therefore are not considered in that analysis of the photoelectron peak. During transport to the nanoparticle surface, the electron might encounter elastic collisions, which will change the electron's initial photoemission angle θ. The differential elastic scattering cross-section defines the probability of the electron to be scattered in a given direction in respect to its initial direction. [34,35] Elastic scattering reduces the molecular (gas-phase) value of β to the value observed in experiment (Fig. 5). The extent of that reduction depends on the number of elastic scattering events. For example, in the limit of infinite elastic scattering events, an isotropic electron distribution would be observed, corresponding to β = 0. In reality, the number of elastic collisions is limited by the value of the IMFP. This assumes that inelastic collisions only decrease the photoelectron's KE, eliminating it from further analysis, and will not affect the photoelectron's angular distribution.

Werner[35] used a Monte Carlo simulation to describe the transport of electrons in solids, which later resulted in a NIST database: Simulation of electron spectra for surface analysis (SESSA).[36] For analysis, the SESSA software uses differential elastic scattering cross-sections calculated by the program developed by Yates.[37] The SESSA software was used in this work to simulate photoelectron spectra and obtain EMFP which correspond to the experimentally measured β parameter. To minimize the number of assumptions, all parameters used as inputs to the SESSA software corresponded to ones used in the actual experiment, including the IMFP values obtained as described above. Simulations started from an arbitrary value of the EMFP to obtain photoelectron spectra at different detection angles. The β parameter was obtained from the angular dependence of the photoelectron peak intensity. The inputted EMFP was varied until the simulated and experimental β parameters matched.

Figure 6a shows the calculated elastic mean free paths at different KE. The values are as small as 0.21 nm for electron KE = 11.8 eV and reaching value of 0.65 nm for electron with KE = 41.7 eV. Figure 6b shows the IMFP/EMFP ratio, which provides information on the average number of elastic scattering events experienced by a photoelectron: The IMFP is the mean distance traveled by a photoelectron before it escapes from a sample and is detected or scatters inelastically. The EMFP is the mean distance between elastic collisions. Therefore the IMFP/EMFP ratio gives mean number of elastic collisions experienced by a photoelectron before its detection or inelastic scattering. The number of elastic interactions decreases with increasing electron KE, ranging from ~8 at 11.8 eV KE to 1.3 for 41.7 eV KE. Antonsson et al. observed a similar decrease in the number of elastic collisions for higher KE electrons in $SiO_2$.[10] A high number of elastic collisions for low KE electrons significantly changes the trajectory of the electron (as observed in the electrons' PAD). Increasing elastic collisions can lead to some cases where the electron approaches the nanoparticle surface at a high impact angle, which can lead to the total internal reflection of the electron as depicted in Fig. 1d.[38,39] Because the total path of the electron is limited by the IMFP, the reflected electron may scatter inelastically before escaping into vacuum. This effect is observed in the experimental photoelectron spectra (Fig. 2b) as the secondary electron signal decreases below 3 eV KE.

### c) Total inelastic and elastic scattering cross-sections

In general, the mean free path $\lambda$ is inversely proportional to the number density of targets $n$ and their total scattering cross-section $\sigma$. In the case of inelastic scattering, the inelastic mean free path $\lambda_{inelastic}$ is given by:

$$\lambda_{inelastic}(hv) = \frac{1}{n\sigma_{inelastic}(hv)}, \quad (5)$$

where $\sigma_{inelastic}(hv)$ is the inelastic scattering cross-section. Using equation (5) it is possible to calculate the inelastic scattering cross-section from our experimental values of the IMFP and number density of carbon atoms (only carbon atoms are considered here due to their higher scattering cross-section compared to hydrogen). The number density calculated from the molar mass and density of liquid squalene is $3.774 \cdot 10^{22}$ cm$^{-3}$. Using eq. 5, squalene's inelastic scattering cross-section per carbon atom $\sigma_{inelastic}(hv)$ dependence is shown in Fig. 7a (black symbols). The obtained inelastic scattering cross-section monotonically increases from $1.6 \cdot 10^{-16}$ cm$^2$ for 11.8 eV KE to $3.8 \cdot 10^{-16}$ cm$^2$ for 51.7 eV KE.

Several processes are responsible for inelastic collisions of electrons in the condensed phase such as dissociative electron attachment, vibrational and electronic excitations, as well as ionization. The most important channel, especially for electrons with KE above ~10 eV, is the ionization process.[40] Because of this, we compare the experimental value for the squalene's inelastic scattering cross-section with the electron impact ionization cross-sections obtained for gas-phase molecules. There are several variations between individual gas-phase molecules and condensed matter,[41,2] affecting electron interactions. One of them is important when interactions of molecules and atoms cannot be neglected. In the case considered here, squalene molecules do not react chemically with each other upon condensation and therefore do not change the electron interaction. The second difference arises due to the quantum-mechanical nature of the electron, when de Broglie wavelength of the electron is comparable with the interatomic distances in a molecule. To include this possibility, the data is compared with hydrocarbon molecules containing different number of carbon atoms.

For comparison, the electron impact ionization cross-section data for similar molecules are shown in Fig. 7a. The data are obtained using the binary-encounter Bethe model, which successfully reproduce experimental data. While electron impact ionization cross-sections for methylene[42] ($CH_2$) and methane[43] ($CH_4$) are shown without any changes, the cross-section for allyl[44] ($C_3H_5$) is scaled by factor of 3 to obtain a cross-section per C atom. All three cross-sections have a threshold around 10 eV, correlating with the ionization energy of the corresponding molecules. Above the threshold, the cross-section increases, reaching a maximum around 100 eV (not shown in Fig. 7a). The cross-sections for methane, methylene, and scaled cross-section of allyl demonstrate similar trend and magnitude in Fig. 7a. The inelastic scattering cross-section obtained for squalene demonstrates a similar trend, which supports our hypothesis that above 10 eV KE, the dominant inelastic scattering process is electron impact ionization.

The elastic scattering cross-sections can be found from the EMFP with a similar approach as was used above and are presented in Fig. 7b. The values are compared with experimental values of elastic scattering cross-section for methane[45] and ethylene.[46] Since the elastic scattering cross-section of squalene is normalized to the number density of C atoms, the cross-section of ethylene is divided by two to scale it to $CH_2$. The literature data have a maximum cross-section at ~10 eV electron KE and decay exponentially with increasing KE. A similar behavior is observed for squalene's elastic scattering cross-section. Moreover, the absolute values of squalene's cross-section are also similar to those of methane and ethylene, providing a good check that there are no systematic errors in our analysis.

The published experimental values of the elastic scattering cross-section for methane and ethylene decrease for KE below 10 eV.[45,46] This implies that the EMFP is increasing for low KE electrons, suggesting that the electron elastic scattering processes might be getting less important at low KE.

**d) Depth profiling using low KE electrons**

The energy dependence of the IMFP represented by the universal curve is mostly used for sample depth profiling during XPS experiments. Increasing the photon energy results in increasing KE of the photoelectrons. When electrons have KE above the universal curve minimum (50 – 100 eV), increasing KE leads to a larger IMFP or a deeper probing depth. The low KE range of the universal curve is less studied and also demonstrates less "universal" character for different materials and therefore is rarely used for depth profiling of the sample surface.

The IMFP of squalene, obtained in this work and shown in Fig. 6a, demonstrates predicable decrease for KE between 10 – 50 eV. The minimum of IMFP at 50 eV KE corresponds to 0.7 nm, whereas the maximum IMFP value corresponds to 1.6 nm at 11.8 eV KE. There is a 2.3 fold difference between the IMFP values. It would be instructive to compare that value to the high KE range of the universal curve, which is normally used for the depth profiling experiments. For example, for electrons penetrating through a guanine layer, a change of KE by factor of 3 (from 500 to 1500 eV) leads to increase of EAL by the similar factor of 2.4 (from 1.3 to 3.1 nm).[47] Experimentally it might be harder to generate such KE difference because of the need for synchrotron beamlines generating X-rays in broad photon energy range.

Nevertheless, for the low KE range studied here, the picture is a bit more complex. For high KE electrons, the EAL is approximately equal to the IMPF, because the effect of elastic scattering is negligible in this energy regime (Fig. 6b, right cartoon). For the low KE electrons the situation dramatically changes, as it was demonstrated above: elastic cross-sections are high, leading to high number of elastic collisions, which may significantly change initial electron directions, essentially decreasing the EAL (Fig. 6b, left cartoon). It is of crucial importance to account for elastic scattering effect when analyzing probing depth of low KE electrons. This may include consideration of number of elastic scattering events (i.e. IMFP/EMFP) as well as directionality of elastic scattering in terms of differential elastic scattering cross-sections.

**CONCLUSIONS**

Soft X-ray photoelectron spectroscopy has been used to probe inelastic and elastic scattering in the condensed phase, represented by the liquid branched hydrocarbon squalene. The VMI images and reconstructed spectra collected above the C 1s edge from the unsupported nanoparticles of squalene provided information on the photoelectron signal intensity, angular distribution and secondary electrons. These data combined allowed for extraction of IMFP directly from the experimental data. The Monte Carlo simulation coupled with the experimental values of PAD and IMFP were used to extract EMFP. While IMFP decreases from 1.6 nm to 0.7 nm for KE between 12-50 eV, the EMFP increases from 0.2 nm to 0.7 nm for KE between 12-40 eV. From the IMFP and EMFP values, corresponding electron scattering cross-sections are determined. Electron impact ionization is the dominant inelastic scattering mechanism in the KE regime measured here. The use of the IMFP for depth profiling XPS experiments could be complicated by strong elastic scattering at low KE. For instance, an average photoelectron with 12 eV KE scatters elastically 8 times before inelastically scattering or escaping to vacuum. The technique developed here has promise for characterization of electron transport parameters in the condensed phase.

**ACKNOWLEDGEMENTS**

This work is supported by the Condensed Phase and Interfacial Molecular Science Program, in the Chemical Sciences Geosciences and Biosciences Division of the Office of Basic Energy Sciences of the U.S. Department of Energy under Contract No. DE-AC02-05CH11231. M.I.J. is supported by a NSF Graduate Research Fellowship under DGE-1106400. This research used resources of the Advanced Light Source, which is a DOE Office of Science User Facility under Contract No. DE-AC02-05CH11231.

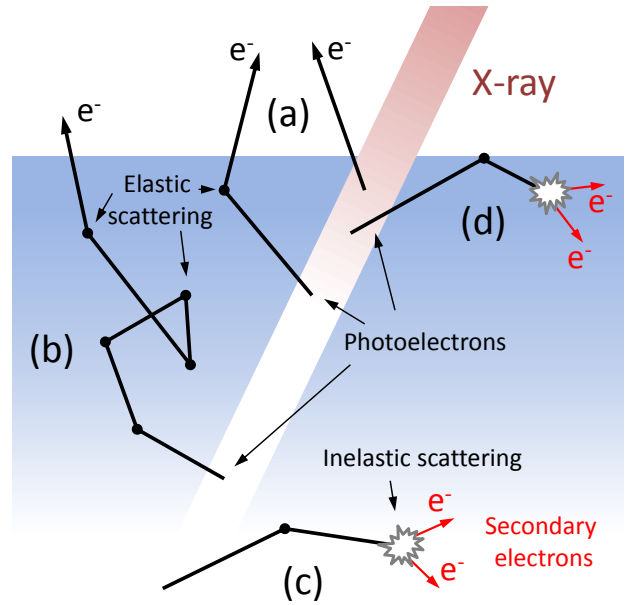

Figure 1. Different processes observed after photoemission of electrons. (a) Weak elastic scattering (elastic scattering represented by black dots). (b) Strong elastic scattering. (c) Inelastic scattering generates two secondary electrons (shown in red). (d) Total internal reflection of an electron is terminated by inelastic scattering.

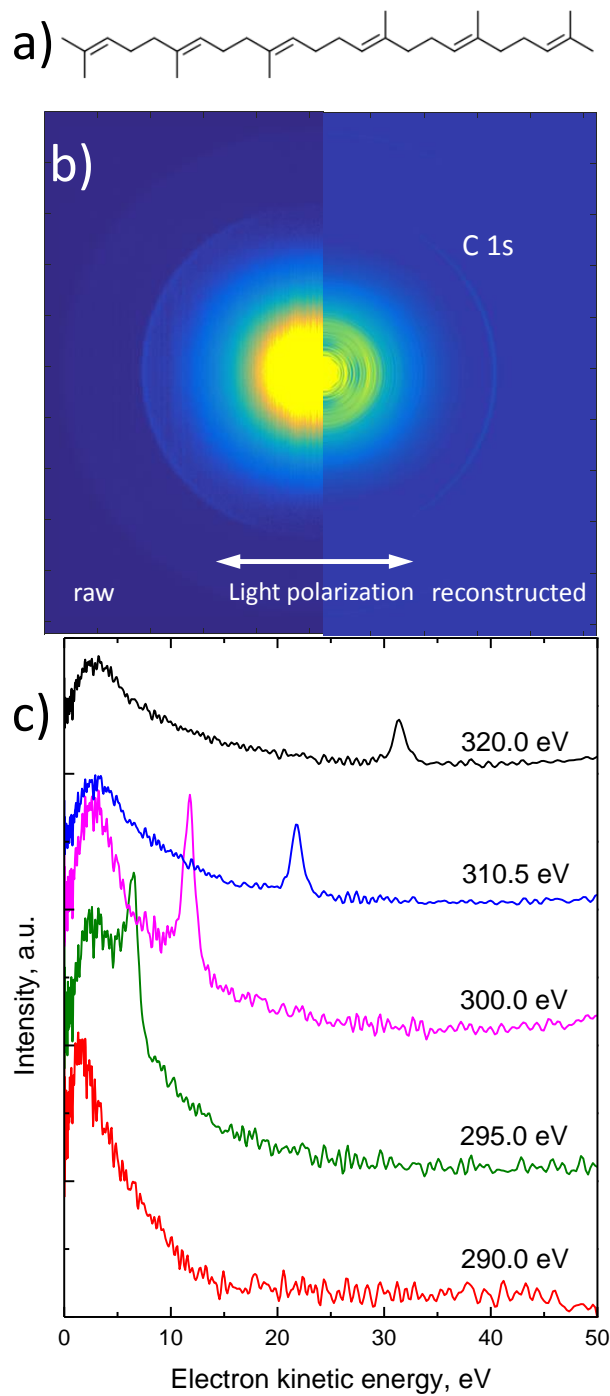

Figure 2. (a) Molecular structure of squalene. (b) Velocity map image collected from squalene nanoparticles irradiated by 315 eV X-ray photons. The left side is the raw image, whereas the right side is the image reconstructed by the pBASEX algorithm. (c) Extracted photoelectron spectra from the VMI images collected at different photon energies.

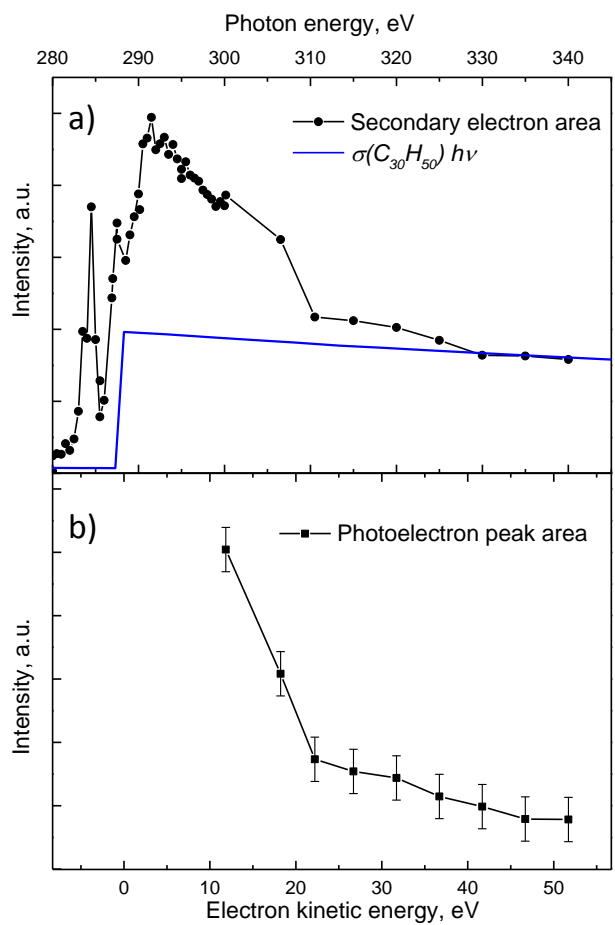

Figure 3. (a) X-ray absorption spectra of squalene. The black line and symbols corresponds to the area of secondary electron background. The blue line represents a product of semi empirical photoabsorption cross-section of $C_{30}H_{50}$, taken from Ref. [19] and photon energy. (b) Area of primary photoelectron signal. All data is normalized to the photon flux.

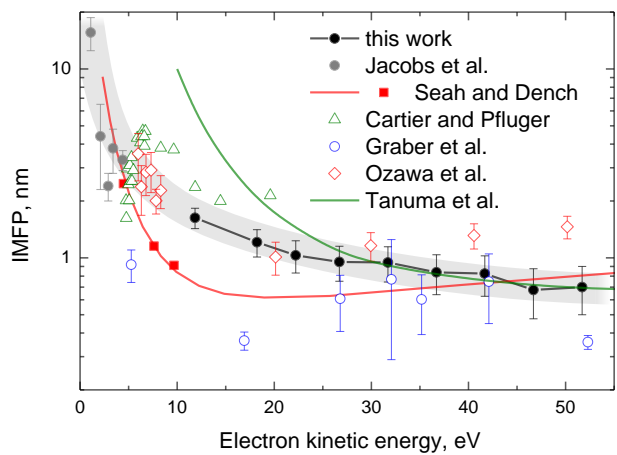

Figure 4. Absolute values of inelastic mean free path for squalene (black circles). Literature values of IMFP or EAL are shown for comparison. Jacobs et al.: EAL for squalane ($C_{30}H_{62}$);[13] Seah and Dench: IMFP for organic compounds;[21] Cartier and Pfluger: EAL for n-$C_{36}H_{74}$ paraffin;[23] Graber et al.: EAL values for PTCDA;[25] Ozawa et al.: EAL value for a π-conjugated organic semiconductor material;[26] Tanuma et al.: IMFP value for 26-n-paraffin.[27] Gray line depicts tentative behavior of squalene's IMFP at low KE.

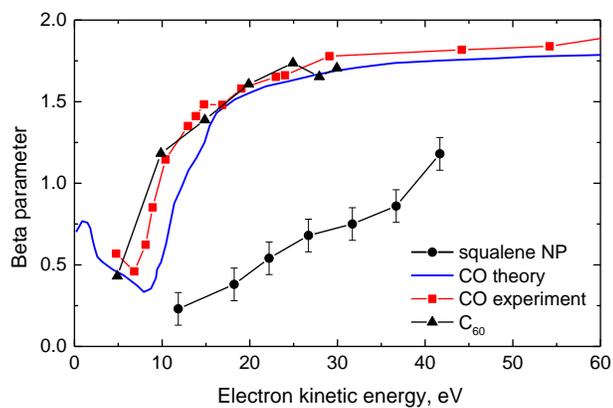

Figure 5. Black circles – beta parameter extracted from experimental VMI spectra of squalene nanoparticles. Blue line and red circles – theoretical[30] and experimental[31] beta parameter data for gas-phase CO. Black triangles – experimental beta parameter for gas-phase $C_{60}$.[32]

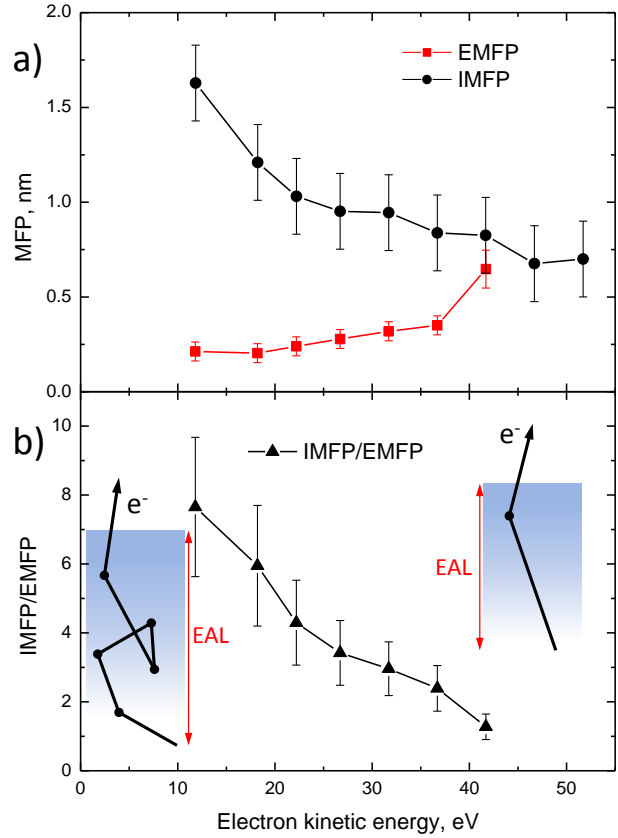

Figure 6. (a) Electron elastic mean free path determined from the experimental data for squalene (red squares) compared to IMFP (black circles). (b) IMFP to EMFP ratio, representing the mean number of elastic collisions per one inelastic collision. Two plausible electron trajectories, corresponding to strong elastic scattering (left) and weak elastic scattering (right) are presented.

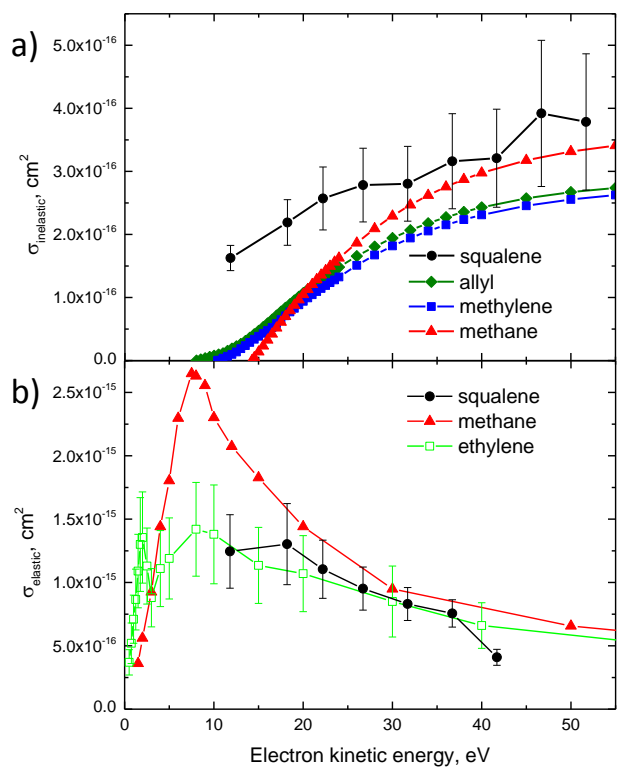

Figure 7. Inelastic and elastic electron scattering cross-sections. (a) Black – inelastic scattering cross-section obtained from the experimental IMFP for squalene, scaled to cross-section per single C atom. For comparison, experimental electron impact ionization cross-sections of allyl (also scaled by factor of 3),[44] methylene,[42] and methane[43] are shown. (b) Electron elastic scattering cross-section, calculated from the EMFP for squalene (black), compared to the experimental values of electron elastic scattering cross-section for methane[45] and ethylene,[46] scaled to $CH_2$.